# A Tale of 3 Dwarf Planets: Ices and Organics on Sedna, Gonggong, and Quaoar from JWST Spectroscopy

J.P. Emery[1], I. Wong[2,3], R. Brunetto[4], J.C. Cook[5], N. Pinilla-Alonso[6], J.A. Stansberry[7,1,8], B.J. Holler[7], W.M. Grundy[8,1], S. Protopapa[9], A.C. Souza-Feliciano[6], E. Fernández-Valenzuela[6], J.I. Lunine[10], D.C. Hines[7].

[1]Northern Arizona University; Flagstaff, Arizona (joshua.emery@nau.edu)
[2]NASA Goddard Space Flight Center; Greenbelt, Maryland
[3]American University; Washington, D.C.
[4]Institut d'Astrophysique Spatiale, Université Paris-Saclay, CNRS; Paris, France
[5]Pinhead Institute; Telluride, CO
[6]Florida Space Institute, University of Central Florida; Orlando, Florida
[7]Space Telescope Science Institute; Baltimore, Maryland
[8]Lowell Observatory; Flagstaff, Arizona
[9]Southwest Research Institute; Boulder, CO
[10]Cornell University; Ithaca, New York

## Abstract

The dwarf planets Sedna, Gonggong, and Quaoar are interesting in being somewhat smaller than the methane-rich bodies of the Kuiper Belt (Pluto, Eris, Makemake), yet large enough to be spherical and to have possibly undergone interior melting and differentiation. They also reside on very different orbits, making them an ideal suite of bodies for untangling effects of size and orbit on present day surface composition. We observed Sedna, Gonggong, and Quaoar with the NIRSpec instrument on the James Webb Space Telescope (JWST). All three bodies were observed in the low-resolution prism mode at wavelengths spanning 0.7 to 5.2 μm. Quaoar was additionally observed at 10x higher spectral resolution from 0.97 to 3.16 μm using medium-resolution gratings. Sedna's spectrum shows a large number of absorption features due to ethane ($C_2H_6$), as well as acetylene ($C_2H_2$), ethylene ($C_2H_4$), $H_2O$, and possibly $CO_2$. Gonggong's spectrum also shows several, but fewer and weaker, ethane features, along with stronger and cleaner $H_2O$ features and $CO_2$ complexed with other molecules. Quaoar's prism spectrum shows even fewer and weaker ethane features, the deepest and cleanest $H_2O$ features, a feature at 3.2 μm possibly due to HCN, and $CO_2$ ice. The higher-resolution medium grating spectrum of Quaoar reveals several overtone and combination bands of ethane and methane ($CH_4$). Spectra of all three objects show steep red spectral slopes and strong, broad absorptions between 2.7 and 3.6 μm indicative of complex organic molecules. The suite of light hydrocarbons and complex organic molecules are interpreted as the products of irradiation of methane. We infer that the differences in apparent abundances of irradiation products among these three similarly-sized bodies are likely due to their distinctive orbits, which lead to different timescales of methane retention and to different charged particle irradiation environments. In all cases, however, the continued presence of light hydrocarbons implies a resupply of methane to the surface. We suggest that these three bodies have undergone internal melting and geochemical evolution similar to the larger dwarf planets and distinct from all smaller KBOs. The feature identification presented in this paper is the first step of analysis, and additional insight into the relative abundances and mixing states of materials on these surfaces will come from future spectral modeling of these data.

## 1. Introduction

The detection and characterization of Kuiper Belt Objects (KBOs; also called Trans-Neptunian Objects, TNOs) ushered in a new era in the understanding of Solar System history. These bodies act as tracers of the gravitational currents that have sculpted the outer Solar System, and their orbits reveal a very dynamic history. The giant planets themselves have migrated in semi-major axis, with Uranus and Neptune in particular having moved outward significantly from their formation regions (e.g., Malhotra et al. 1993, Gomes et al. 2005), and it is possible that there was a fifth giant planet that was lost (e.g., Nesvorný 2011). This planetary migration placed the KBOs onto their present-day orbits and scrambled the original compositional gradients that existed in the solar nebula (with the exception of cold classical KBOs, which are thought to remain roughly where they formed).

Compositions of sub-planet sized bodies (e.g, asteroids, comets, Kuiper Belt Objects) provide a window into the conditions of the solar nebula in the regions in which they formed, provided that any alterations to the surface from exposure to the space environment or other evolutionary events (including differentiation for the largest of these bodies) can be understood. The significant gravitational scattering and mixing of objects from all parts of the early outer Solar System complicates the interpretation of nebular conditions in that region. Because the dynamical interactions that led to the present day Kuiper Belt are stochastic (e.g., Tsiganis et al. 2005), a deterministic reconstruction of the origin locations of KBOs is not generally possible. However, compositions themselves may provide clues that will enable mapping of those original locations and, thereby, provide important constraints on dynamical models of exactly how the scattering occurred.

Spectral analyses, coupled with thermodynamic models of volatile loss, have revealed that the largest KBOs (i.e., the dwarf planets Pluto, Eris, Haumea, and Makemake) are sufficiently massive to retain extremely volatile compounds (e.g., $CH_4$, $N_2$, CO; e.g., Schaller and Brown 2007b, Johnson et al. 2015). They therefore comprise a special class of objects in which atmospheric processes and/or volatile migration heavily influence surface compositions, refreshing their surfaces (Stern and Trafton 2008, Young et al. 2020). Much smaller bodies lose these volatiles to space (Schaller and Brown 2007b, Johnson et al. 2015, Lilly et al. 2020, Lisse et al. 2021, 2022, Davidsson et al. 2021, Pahri & Prialnik 2023). Prior to observations by the James Webb Space Telescope (JWST), the surfaces of these smaller KBOs appeared to mostly be mixtures of $H_2O$ ice and dark, reddish material (e.g., Barucci et al. 2007, Brown et al. 2012), with $CH_3OH$ also detected on a few bodies (Cruikshank et al. 1998, Barucci et al. 2006, Grundy et al. 2019). With the extended wavelength range and improved sensitivity of JWST, other non-volatile and volatile ices (e.g., $CO_2$, CO) have been detected (De Prá et al. 2024, Brown and Fraser 2023, Pinilla-Alonso et al. 2024). For those small bodies, the detailed composition(s) of the dark material(s) is important for understanding original compositions and subsequent processing.

Several KBOs have sizes between these extremes, with diameters around 1000 km, including (50000) Quaoar, (225088) Gonggong, and the inner Oort Cloud (IOC) object (90377) Sedna. These bodies should hold onto original volatile ices longer than the smaller objects while (likely) not supporting atmospheres that could partially block incoming UV or high energy particle radiation or undergoing surface refreshing processes of the larger dwarf planets (e.g., Johnson et al. 2015, Young et al. 2020). They therefore hold the promise of revealing details of the long-term radiation processing of volatile ices that is masked by ongoing seasonal/atmospheric processes on the larger dwarf planets.

(90377) Sedna (2003 VB12) is the largest (D~995 km; Pál et al. 2012) of the three known members of the Inner Oort Cloud (IOC) population (Brown et al. 2004, Sheppard et al. 2019). With a perihelion distance of 76 AU and aphelion distance of 937 AU, Sedna spends most of its orbit outside of the heliosphere, subjecting it to a different irradiation environment than most KBOs. Ground-based visible and near-infrared (NIR; 0.4 to 2.5 μm) spectra of Sedna show an absorption band near 2.3 μm, hints of $H_2O$ ice, and a steep red spectral slope and visible and near-infrared wavelengths suggestive of complex organics (e.g., Barucci et al. 2005, 2010). These previous studies have attributed the 2.3-μm absorption to $CH_4$ and/or $CH_3OH$ ice. Photometry at 3.6 and 4.5 μm measured with the Infrared Array Camera (IRAC) on the Spitzer Space Telescope indicated strong absorption bands relative to the 2-μm region and supported spectral models that include $H_2O$ ice and simple hydrocarbons (Emery et al. 2007). However, the low signal to noise ratio (S/N) of the NIR spectra and photometric nature of the IRAC data hinder detailed compositional analysis. With equilibrium temperatures of ~30 K and ~9 K at perihelion (~76 AU) and aphelion (~937 AU), respectively, a wide range of ices should be stable on the surface.

(22508) Gonggong (2007 OR10) is the fifth largest known KBO (D~1,230 km; Kiss et al. 2019), after Pluto, Eris, Haumea, and Makemake. Its eccentric and inclined orbit originally garnered a scattered disk classification, but a more recent analysis[1] (Gladman and Volk 2021) indicates that Gonggong is in the 3:10 mean motion resonance with Neptune. Ground-based visible and near-infrared observations show a steep red spectral slope and $H_2O$ ice absorptions at 1.5 and 2.0 μm (Brown et al. 2011a). Those data do not have sufficient S/N to search for $CH_4$ or other hydrocarbons, but Brown et al. (2011a) argue that, based on its size, Gonggong may retain $CH_4$ on the surface and irradiation of that $CH_4$ could be responsible for the steep red spectral slope. Johnson et al. (2015) point out that with its eccentric orbit, Gonggong may support an atmosphere during part of its 554-yr orbit that freezes out at $r_h \gtrsim 70$ AU. Gonggong has a moon, Xiangliu, from whose orbit a density of ~1.75 g cm$^{-3}$ has been determined (Kiss et al. 2017, 2019).

(50000) Quaoar (2002 LM60) has a diameter between those of Sedna and Gonggong (1,086 km; Periera et al. 2023). It has a nearly circular orbit around the Sun at ~44 AU with a moderate inclination of ~8°, placing it among the Hot Classical dynamical class of KBOs. Crystalline $H_2O$ is clearly visible in ground-based NIR spectra, and $CH_4$ and $C_2H_6$ (ethane) absorption features have also been detected (Jewitt and Luu 2004, Schaller and Brown 2007a, Dalle Ore et al. 2009, Barucci et al. 2015). Barucci et al. (2015) suggest the presence of $N_2$ based on the wavelength position of the 1.65-μm feature (assuming it is strongly influenced by an overlapping $CH_4$ band) and $NH_3 \cdot OH$ to improve the model fits in the 2.0 to 2.3-μm range. Like Sedna and Gonggong, Quaoar has a steep red visible to near-infrared (VNIR) spectral slope, likely due to complex organic molecules (tholins) that formed from irradiation of ices (Marchi et al. 2003, Dalle Ore et al. 2009, Barucci et al 2015). Stellar occultations show no sign of an atmosphere, placing upper limits of ~10 to 20 nbar on any atmosphere (Braga-Ribas et al. 2013, Arimatsu et al. 2019), but do reveal two rings around Quaoar (Morgado et al. 2023, Pereira et al. 2023). Quaoar's density of ~2.0 g cm$^{-3}$ (Pereira et al. 2023, Fraser et al. 2013), computed from the orbit of its moon Weywot, is consistent with those of the larger dwarf planets.

---

[1] https://web.archive.org/web/20190524032543/https://www.boulder.swri.edu/~buie/kbo/astrom/225088.html

Previous observations show similar spectral properties among these three KBOs that have similar sizes but very different orbits (Table 1). They all exhibit very red spectral slopes, $H_2O$ ice absorption bands, and hints of absorption near 2.3 μm that has been attributed to light hydrocarbons such as $CH_4$ (Barucci et al. 2005, Dalle Ore et al. 2009, Brown et al. 2011a). A reasonable hypothesis to explain these similarities is that all three bodies are large enough to retain $CH_4$ ice, irradiation of which leads to the red coloration. In other words, the present-day surface composition depends mostly on the size, not on the current orbits. Alternatively, recognizing that the earlier ground-based spectra, particularly of Gonggong and Sedna, have relatively low S/N, it is possible that the surface compositions are different, and that either the current orbits are important in the surface evolution or that the original compositions of these bodies were distinct. Spectra of higher S/N and that extend beyond 2.5 μm are best suited to distinguish among these hypotheses, and we present such new spectra here.

**Table 1.** Target Properties

| Object | Diam (km) | $p_v$ | Moon | a (AU) | e | i (°) | q (au) | Q (au) | Dyn. Class[a] |
|---|---|---|---|---|---|---|---|---|---|
| Sedna | 995[b] | 0.32 | --- | 506 | 0.85 | 11.9 | 76.2 | 937 | IOC |
| Gonggong | 1,230[c] | 0.14 | Xiangliu | 67.5 | 0.50 | 30.6 | 33.8 | 101.2 | Res. |
| Quaoar | 1,086[d] | 0.12 | Weywot | 43.7 | 0.04 | 8.0 | 41.9 | 45.5 | HC |

[a]IOC=Inner Oort Cloud, Res=Resonant (3:10), HC=Hot Classical. [b]Pal et al. (2012). [c]Kiss et al. (2019). [d]Pereira et al. (2023)

## 2. Observations and data reduction

The spectra of Sedna, Gonggong, and Quaoar were measured using the integral field unit (IFU) of the Near-Infrared Spectrograph (NIRSpec) on the James Webb Space Telescope (JWST). The observations were obtained as part of Guaranteed Time Observations Programs 1191 (Gonggong), 1272 (Sedna), and 1273 (Quaoar). All three targets were observed with the low spectral resolution PRISM grating, which provides continuous wavelength coverage from 0.7 to 5.3 μm with a spectral resolving power, R=λ/Δλ, ranging from 30 to 300. Additionally, Quaoar was observed with the medium spectral resolution (R ~ 1000) grating/filter combinations G140M/F100LP (0.97-1.89 μm) and G235M/F140LP (1.66-3.17 μm). A pair of dithered exposures was obtained at each grating setting for all three targets. Additional details of the observations are provided in Table 2.

**Table 2.** Observation Details

| Object | Grating/Filter | Date (UT) | Start Time (UT) | $t_{int}^{a}$ (s) | r (au) | Phase angle (°) | Prog/Obs[b] numbers |
|---|---|---|---|---|---|---|---|
| Sedna | PRISM/Clear | 13 Sep 2022 | 04:53 | 759 | 83.90 | 0.65 | 1272/8 |
| Gonggong | PRISM/Clear | 4 Nov 2022 | 20:24 | 905 | 89.02 | 0.60 | 1191/15 |
| Quaoar | G140M/F100LP | 1 Sep 2022 | 13:07 | 613 | 42.74 | 1.22 | 1273/4 |
| | G235M/F170LP | 1 Sep 2022 | 13:26 | 1780 | 42.74 | 1.22 | 1273/4 |
| | PRISM/Clear | 1 Sep 2022 | 14:06 | 934 | 42.74 | 1.22 | 1273/4 |

[a]Total integration time across the two dithered exposures.
[b] JWST program and observation numbers

The data processing and spectral extraction methodology was largely identical to that used in other recent studies of KBOs observed with the NIRSpec IFU (see Grundy et al. 2023 for a detailed description). The fully calibrated spectral data cubes, containing stacks of 2D sky-projected wavelength slices with a 3''x3'' field of view and pixel scale of 0.1", were constructed by running the raw uncalibrated data files through the first two stages of Version 1.11.1 of the official JWST pipeline (Bushouse et al. 2022), with all relevant calibration reference files drawn from context jwst_1100.pmap of the JWST Calibration Reference Data System. In between the two stages, the residual correlated read-noise in each column of the 2D detector countrate images was manually removed by subtracting the 200-pixel-wide moving median along the columns, calculated exclusively over the off-sky pixels that are situated between and around the dispersed IFU spectral slices. The irradiance spectra were extracted using a specialized PSF fitting procedure. The local template PSF was constructed at each wavelength slice of the IFU data cube by median-averaging the 10 adjacent slices in both directions, subtracting the background level, and normalizing the template to a unit sum. The local template PSF was then fit to the original wavelength slice with a scaling factor and a constant background level, with the resultant best-fit scaling factor taken as the corresponding irradiance value at that wavelength in the spectrum. This process was applied to each dithered exposure individually, with the final combined spectrum derived by cleaning 3σ outliers in the individual spectra using a 20-pixel-wide moving median filter and averaging each dither pair together.

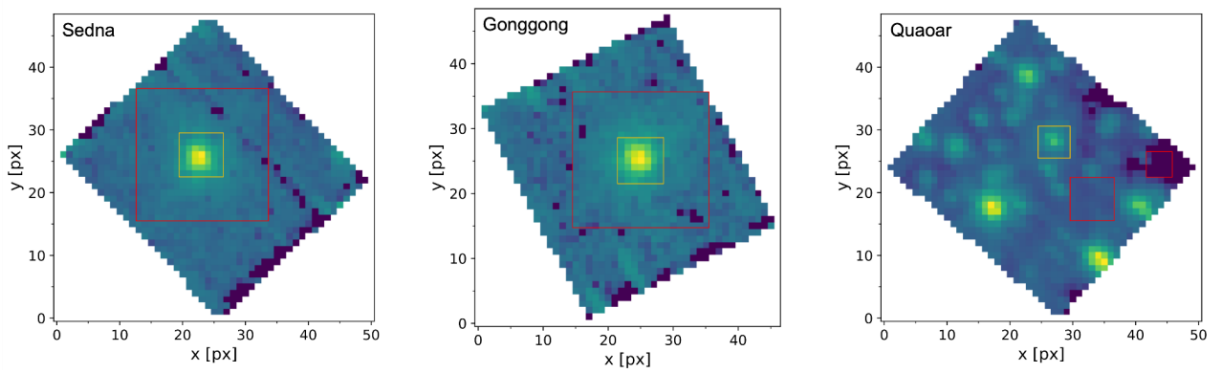

**Figure 1.** IFU images from one of the two PRISM grating observations of Sedna, Gonggong, and Quaoar, collapsed along the entire wavelength axis. For Sedna and Gonggong, the 7x7 pixel yellow box

denotes the spectral extraction region, while the 21x21 pixel red box marks the inner boundary of the background region. Quaoar lies in a dense field of background stars, and all pixels outside of the 5x5 pixel yellow extraction box were masked, except for the two designated background regions, marked in red, which are free from visible astronomical sources.

Fig. 1 shows representative wavelength-collapsed IFU slices from the PRISM grating observations of Sedna, Gonggong, and Quaoar. The fields of view around Sedna and Gonggong were free of other astronomical sources. The satellites of Gonggong and Quaoar are too small and faint to be detected in these data. For these observations, the background region (used when constructing and fitting the template PSFs) was defined as all pixels outside of a 21x21 pixel box, centered on the target's centroid position. The template PSF was fixed to zero in the background region. Previous work has demonstrated that relatively small spectral extraction apertures that primarily sample the high signal-to-noise central regions of the target PSF produce higher-quality spectra with lower scatter than larger apertures that encompass the extended diffraction peaks at wide separations. A 7x7 pixel spectral extraction box was chosen for Sedna and Gonggong. Pixels located between the spectral extraction and background regions were masked and not considered when constructing and fitting the template PSFs. Spectral extraction of Quaoar required more care, because the target was located in a dense field of background sources. To avoid as much contamination as possible, a smaller 5x5 pixel spectral extraction box was used to extract Quaoar's spectrum. Because of the crowded field, two small apertures free from contamination by background sources were used to measure sky background for Quaoar. In Fig. 1, the spectral extraction and sky background regions are annotated for each observation.

To obtain the reflectance spectra, the irradiance spectra were divided by analogously-extracted spectra of two G-type solar analog stars – SNAP-2 (PRISM grating; Program 1128) and P330E (M gratings; Program 1538). The aperture size in the stellar spectrum extraction was chosen to match the corresponding size used for the KBO target observation. In addition to dividing away the approximate solar spectrum from the irradiance spectra, this process self-consistently removed any instrumental systematics common to both target and stellar observations and corrected for the flux biases inherent in the irradiance spectra due to the wavelength-dependent fraction of the source PSF that falls outside of the fixed spectral extraction box. The normalized JWST reflectance spectra of Sedna, Gonggong, and Quaoar are shown in Fig. 2.

## 3. Results

The JWST spectra generally confirm the spectral shapes and absorptions detected in previous ground-based spectra and Spitzer/IRAC photometry (Barucci et al. 2005, 2010, Emery et al. 2007, Brown et al. 2011a, Marchi et al. 2003, Jewitt and Luu 2004, Schaller and Brown 2007a, Dalle Ore et al. 2009, Brown et al. 2011a, Barucci et al. 2015).  The much higher S/N of the JWST data reveal details of the absorptions, particularly in the 2.2–2.5 µm range, that enable more precise interpretations.  The addition of high-quality spectra at $\lambda > 2.5$ µm makes it possible to search for strong fundamental absorptions and identify materials not detected at shorter wavelengths.

All three objects show notable similarities in their spectra, including:  a steep, red spectral slope at $\lambda < 1.3$ µm, absorptions near 1.5 and 2.0 µm and a curvature in the 4- to 5-µm region all attributable to $H_2O$, spectral structure in the 2.3- to 2.5-µm wavelength range, a broad, deep absorption from ~2.7 to ~3.6 µm, discrete absorptions at 3.4 and 3.5 µm, and a feature near 4.25 µm. There are, however, important differences in the spectra of these three dwarf planets. Quaoar has the sharpest $H_2O$ bands and weaker features in the 2.3–2.5 µm range. Sedna's 1.5 and 2.0 µm bands are distorted relative to pure $H_2O$, suggesting some contribution from other molecules. Sedna also shows an extra absorption not seen in the spectra of Quaoar or Gonggong overprinting the 1.65 µm crystalline $H_2O$ band. Sedna's 2.3- to 2.5-µm features are the deepest of the three objects, and it also shows a larger number of discrete absorption features at $\lambda > 2.5$ µm.

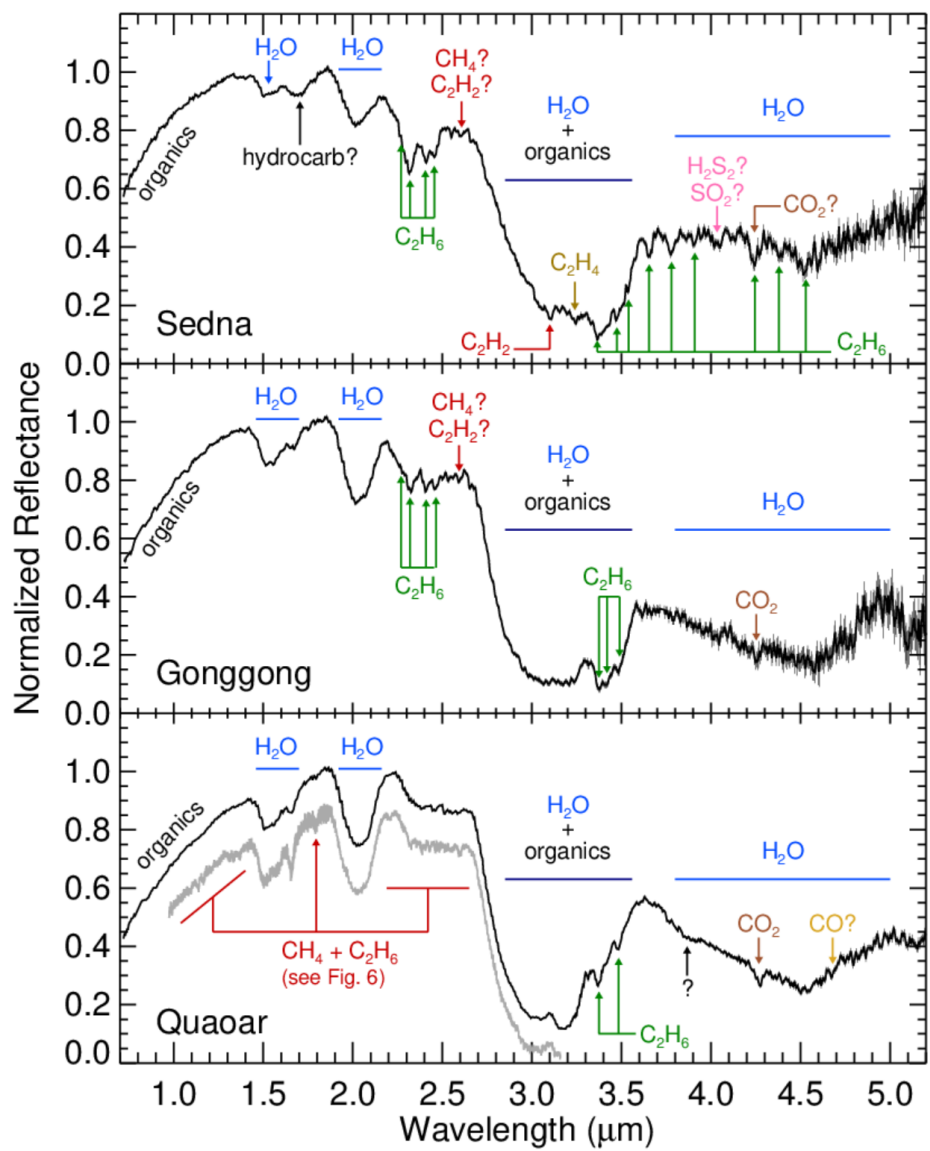

**Figure 2.** JWST NIRSpec spectra of Sedna, Gonggong, and Quaoar. All spectra are normalized to the mean between 1.8 and 1.9 µm. The three dwarf planets share many spectral similarities, but they also differ in several interesting and important ways. The higher spectral resolution medium grating spectrum (shown in gray) of Quaoar resolves features not detected in the prism spectrum, but otherwise agrees very well with the prism spectrum. Details of features detected in the medium grating spectrum are shown in Fig. 6. Section 4 in the text describes the interpretations labeled for the different features.

## 4. Analysis

We focus our analysis here on a description of absorption features in each spectrum and association with specific molecules, where possible. For each object, we provide a table of the band parameters for each of the features detected. The band parameters are computed using Gaussian fits to each band, after division by a local continuum. The continuum is a simple linear fit to regions immediately adjacent to each band. More robust continuum analysis would affect resulting band parameters. The values listed are averages of 10,000 Monte Carlo fits drawing randomly from normal distributions scaled to the uncertainties of each spectral channel in the fit. The listed band parameter uncertainties are the standard deviations from these Monte Carlo fits. These parameter uncertainties do not take into account uncertainties in the shape or slope of the continuum nor the possibility of non-Gaussian shapes or blended bands and therefore may somewhat underestimate the true uncertainty.

The JWST spectra are very rich, and detailed spectral modeling holds the promise of determining abundances of different materials and surface structures (e.g., particle sizes, layering, multiple terrains). With such high-quality spectra containing mixtures of different materials, spectral modeling becomes complex. We have begun spectral modeling (Cook et al. 2023), but the results are not yet finalized. We therefore hold the presentation of spectral modeling for future work.

### 4.1 Sedna

Perhaps the most notable discovery from the JWST spectrum of Sedna is the prevalence of ethane ($C_2H_6$) on the surface. A dozen features in the spectrum can be identified with $C_2H_6$ (Fig. 3, Table 3). Many of these features are at $\lambda > 2.5$ μm, including the strong complex of fundamental stretching bands at ~3.35–3.5 μm. $C_2H_6$ also matches several of the individual bands in the 2.25–2.6 μm absorption complex.

**Table 3.**
Band parameters and assignments for features identified in the spectrum of Sedna.
See text (Section 4) for details about calculations and uncertainties.

| $\lambda_c$ (μm) | | | FWHM (μm) | | | Band Depth (% of local cont.) | | | Assignment |
|---|---|---|---|---|---|---|---|---|---|
| ~1.5 | | | | | | | | | $H_2O$ |
| 1.705 | ± | 0.002 | 0.119 | ± | 0.008 | 5.6 | ± | 0.3 | (hydrocarbons?) |
| 2.020[a] | ± | 0.001 | 0.160 | ± | 0.003 | 15.9 | ± | 0.2 | $H_2O$ |
| 2.310[a] | ± | 0.001 | 0.073 | ± | 0.004 | 18.3 | ± | 0.8 | $C_2H_6$ (+$C_2H_4$?) |
| 2.408 | ± | 0.001 | 0.024 | ± | 0.003 | 6.1 | ± | 0.6 | $C_2H_6$ |
| 2.459 | ± | 0.001 | 0.034 | ± | 0.003 | 6.8 | ± | 0.5 | $C_2H_6$ (+$C_2H_2$?) |
| 2.608 | ± | 0.004 | 0.007 | ± | 0.029 | 2.9 | ± | 0.6 | ($C_2H_2$ or $CH_4$?) |
| 3.102[c] | ± | 0.001 | 0.027 | ± | 0.003 | 21.2 | ± | 2.0 | $C_2H_2$ |
| 3.241[c] | ± | 0.003 | 0.035 | ± | 0.011 | 17.5 | ± | 2.5 | $C_2H_4$ |
| 3.376[a,c] | ± | 0.002 | 0.072 | ± | 0.012 | 44.3 | ± | 4.8 | $C_2H_6$ (+$C_2H_4$?) |
| 3.475[c] | ± | 0.002 | 0.019 | ± | 0.005 | 21.4 | ± | 3.4 | $C_2H_6$ |
| 3.541[b,c] | ± | 0.001 | 0.009 | ± | 0.012 | 14.6 | ± | 2.4 | $C_2H_6$ |

| | | | | | | |
|---|---|---|---|---|---|---|
| 3.656 | ± | 0.001 | 0.032 | ± 0.003 | 17.3 ± 1.2 | $C_2H_6$ |
| 3.777 | ± | 0.002 | 0.045 | ± 0.008 | 15.0 ± 1.1 | $C_2H_6$ |
| 3.906 | ± | 0.002 | 0.013 | ± 0.018 | 9.5 ± 2.0 | $C_2H_6$ |
| 4.034 | ± | 0.004 | 0.053 | ± 0.007 | 10.5 ± 1.4 | ($H_2S_2$? or $SO_2$?) |
| 4.243 | ± | 0.002 | 0.055 | ± 0.008 | 21.5 ± 1.9 | $C_2H_6$ (+$CO_2$?) |
| 4.380 | ± | 0.007 | 0.026 | ± 0.026 | 11.9 ± 3.3 | $C_2H_6$ |
| 4.529 | ± | 0.011 | 0.090 | ± 0.150 | 17.6 ± 3.8 | $C_2H_6$ |

[a]These bands are asymmetric, and the Gaussian fits may not precisely represent the true center.
[b]This band is on a steeply sloped continuum, and the fit may not be precisely reliable.
[c]These bands are within the deep and broad 2.7 to 3.6-μm feature. The FWHM and Band Depth should be taken with consideration that only a local continuum was used.

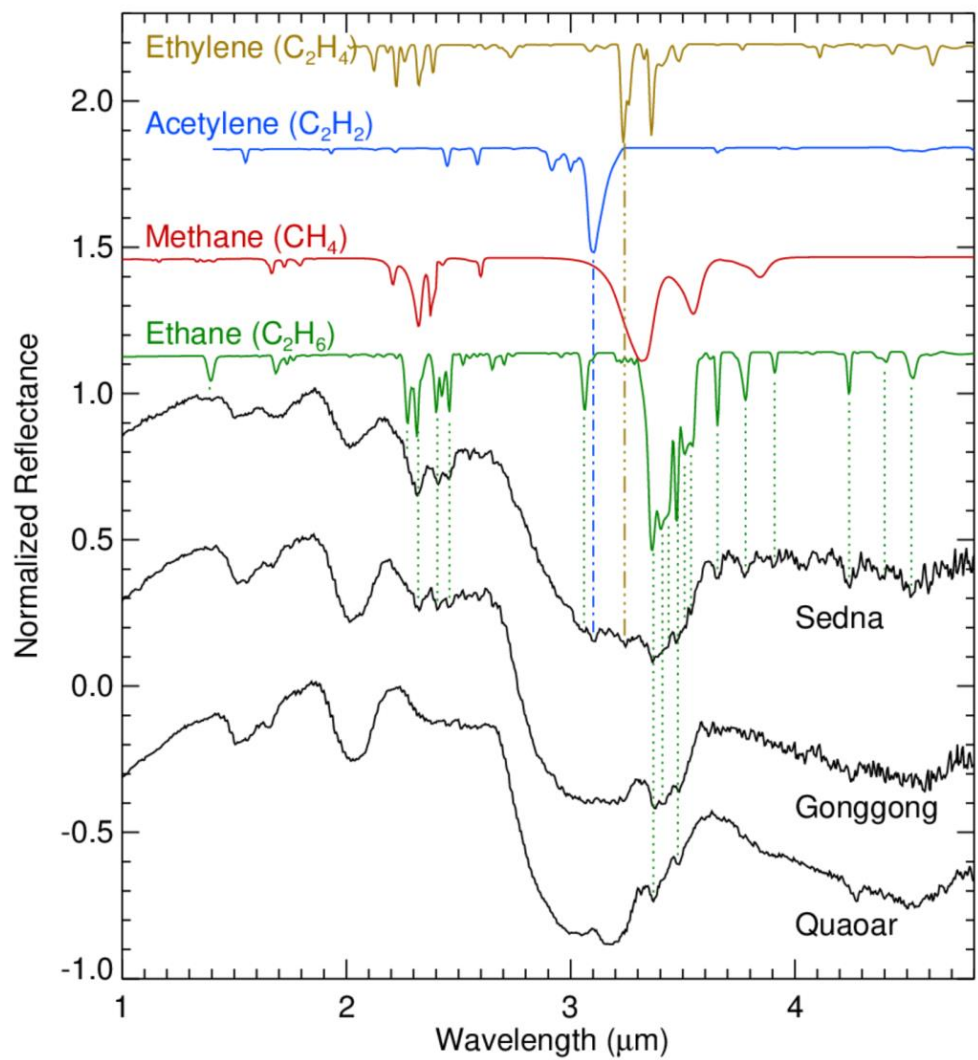

**Figure 3.** The prism spectra of Sedna, Gonggong (shifted -0.5), and Quaoar (shifted -1.0) compared with spectra of ethane, methane, acetylene, and ethylene. Dotted vertical lines indicate locations where ethane bands match features in the measured spectra. The dash-dot vertical line marks the location of the strongest band in the spectrum of acetylene, highlighting that it matches with a feature in the spectrum of Sedna. Similarly, the dash-dot-dot-dot line marks the strongest band in ethylene and its match to a feature in the spectrum of Sedna. Other ethylene bands mostly overlap with ethane features, so it is difficult to assess their potential contributions without detailed spectral modeling. The strong methane fundamental

bands are absent from the spectra of Sedna, Gonggong, and Quaoar. Ethane, acetylene, and ethylene are irradiation products of methane. See the text for additional details. The comparison spectra of ethane, methane, acetylene, and ethylene are computed from optical constants (in particulate mixtures with 15, 20, 10, and 10% amorphous carbon and grain sizes 7.5, 20, 5, and 7.5 μm, respectively) using Hapke theory. For these model spectra, the optical constants are convolved to JWST prism resolution before computing the reflectance spectrum. Sources of optical constants are: acetylene - Hudson et al. (2014a); ethylene - Hudson et al. (2014b); methane - Grundy et al. (2002); ethane - R. Mastrapa (personal communication) and Molpeceres et al. (2016).

Interestingly, evidence for methane ($CH_4$) in the spectrum is much less clear. The strong $CH_4$ bands at ~3.3 and ~3.54 μm are not present in the Sedna spectrum (Fig 3). These absorptions occur in the middle of the strong, broad 2.7- to 3.6-μm absorption, but $C_2H_6$ features are visible within this absorption, so any undetected $CH_4$ must be much less abundant than $C_2H_6$. At shorter wavelengths, the $CH_4$ band at ~2.31 μm overlaps with a $C_2H_6$ band. The two $CH_4$ bands at 2.21 and 2.39 μm do not have corresponding features in the Sedna spectrum, though there are small depressions at those wavelengths that might suggest small amounts of $CH_4$. The 2.60-μm $CH_4$ band does line up with a feature at 2.601 μm in the JWST spectrum that is not explained by $C_2H_6$. Acetylene ($C_2H_2$) also has a band near this position, but at a slightly shorter wavelength of 2.587 μm. The complex of combination and overtone bands in $CH_4$ at ~1.65–1.8 μm and ~1.3–1.4 μm also correspond to small depressions in the Sedna spectrum. If small amounts of $CH_4$ are present at $\lambda$<2.5 μm, but absent at longer wavelengths, it might indicate compositional layering, where $CH_4$ is absent from the topmost surface, but present at smaller amounts slightly deeper. The presence of such layers would be consistent with the conversion of $CH_4$ at the surface to longer chain hydrocarbons due to irradiation and the possibility that Sedna can retain volatiles due to its extreme orbit.

A feature at 3.10 μm is not matched by any $C_2H_6$ or $CH_4$ band. Acetylene ($C_2H_2$) has a strong absorption band that matches this feature (Fig. 3). Acetylene also has bands at 2.453 and 2.587 μm that could contribute to features in the Sedna spectrum (though $C_2H_6$ also has a band at 2.46 μm). Acetylene is a common irradiation product of $CH_4$ (Bennett et al. 2006), so would be expected if the $C_2H_6$ were also produced by irradiation of $CH_4$. Ethylene ($C_2H_4$) is also formed from the irradiation of $CH_4$. A feature at 3.24 μm in the Sedna spectrum is best explained by ethylene. Other bands in the spectrum of ethylene also align with some of the features in the Sedna spectrum, but they often also overlap with $C_2H_6$ (Fig. 3). In addition to the light hydrocarbon ices, complex organic molecules are indicated by the steep red spectral slope and also likely contribute to the broad absorption complex between 2.7 and 3.6 μm.

The Sedna spectrum has a feature at 4.243 μm, close to the location of the $CO_2$ fundamental absorption. The maximum depth of the feature in Sedna's spectrum is at a slightly shorter wavelength than the center of the pure $CO_2$ ice band (Fig. 4). The 2.6- and 2.7-μm overtone and combination bands of $CO_2$ that are detected on other KBOs (De Prá et al. 2024) are not detected in the spectrum of Sedna. $C_2H_6$ has a band that aligns better with the Sedna 4.243-μm feature, but the feature in the Sedna spectrum seems to be deeper and wider relative to surrounding features than the $C_2H_6$ band is to the corresponding surrounding bands in its spectrum. It is possible that there is a small contribution from $CO_2$, which appears to be widespread in the Kuiper Belt (De Prá et al. 2024), but, again, spectral modeling will be required to assess the presence of $CO_2$ more certainly.

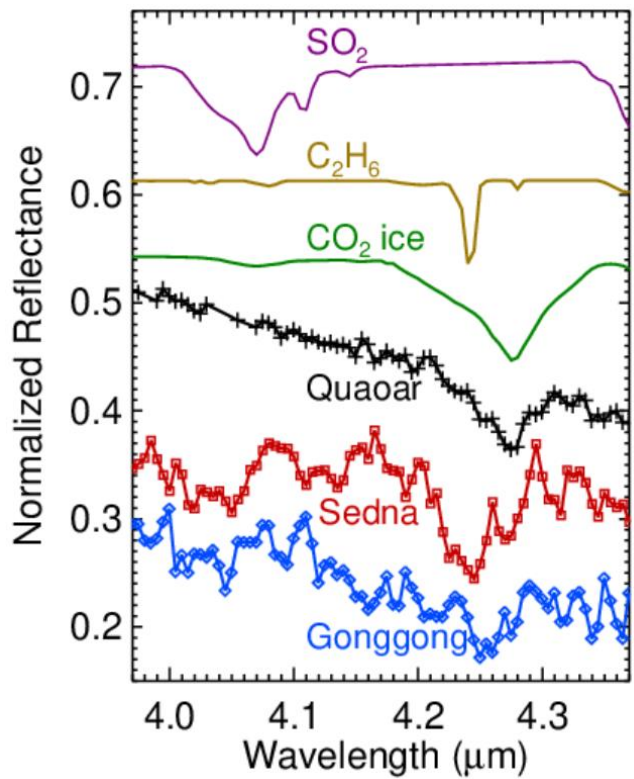

**Figure 4.** The spectra of Sedna, Quaoar, and Gonggong compared to $SO_2$, ethane ($C_2H_6$) and $CO_2$ ice in the ~4.0- to 4.3-μm region. The feature in the spectrum of Quaoar at 4.268 μm is consistent with $CO_2$ ice. The feature at 4.243 μm in the spectrum of Sedna is well aligned with $C_2H_6$. However, it is broader than the $C_2H_6$ band, and deeper than expected relative to surrounding $C_2H_6$ features in the Sedna spectrum. The feature in the spectrum of Gonggong does not fit with either $C_2H_6$ or $CO_2$ ice and may instead be due to $CO_2$ complexed with another material. A feature from ~4.00 to 4.08 μm (centered at ~4.03 μm) in the spectrum of Sedna is close to an $SO_2$ band, but not close enough to be identified as $SO_2$. $H_2S_2$ (not shown) is another candidate for this feature (see text). The spectra of $SO_2$, ethane, and $CO_2$ are computed from optical constants (in a particulate mixtures with 15, 10, and 10% amorphous carbon, respectively) using Hapke theory. Sources of optical constants are: $SO_2$ - Schmitt et al. (1994); $CO_2$ - Hansen (1997, 2005); ethane - see Fig. 1 caption.

A feature at 4.03 μm in the Sedna spectrum remains unidentified. The feature is somewhat close to the $\nu_1+\nu_3$ $SO_2$ band (Salama et al. 1990), but the center of the $SO_2$ band is at slightly longer wavelengths (~4.07 μm) than the center of the Sedna feature. Moore et al. (2007) indicate that this $SO_2$ band can be shifted to shorter wavelengths (~4.05 μm) in mixtures with $H_2O$; perhaps we are seeing a shifted $SO_2$ band. Hydrogen persulfide ($H_2S_2$) has a band at 4.02 μm. $H_2S_2$ is created in irradiation of pure $H_2S$ and in mixtures of $H_2S$ in $H_2O$ and CO (Moore et al. 2007, Garozzo et al. 2010). Garozzo et al. (2010) document cases of complete destruction of $H_2S$ correlated with growth of $H_2S_2$, though they also note cases where the $H_2S_2$ is depleted with additional irradiation. Since we do not see any features due to $H_2S$, if the 4.03-μm feature in Sedna's spectrum is due to $H_2S_2$, it would suggest depletion of the original $H_2S$ in the creation of $H_2S_2$. Laboratory measurements of reflectance spectra and/or optical constants would help test the possible attribution of this feature to $H_2S_2$. As such, the 4.05 µm feature is still to be identified, but $SO_2$ and $H_2S_2$ are viable candidates to explain it.

As mentioned in Section 3, the 1.5 and 2.0 µm features in Sedna's spectrum are in the right location for $H_2O$, but the shapes are somewhat distorted from pure $H_2O$ ice (Fig. 5). The distortion could indicate contaminates (e.g., complex organics) within the ice. Alternatively, the absorption could be caused by some hydrate that contains $H_2O$ molecules, analogous to the distorted bands observed in spectra of Europa and Ganymede (e.g., McCord et al. 1998, McCord et al. 2001) or OH-bearing complex organic molecules. The broad 2.7–3.6 µm absorption in the Sedna spectrum is also not well matched to the shape of the $H_2O$ spectrum. Part of the $H_2O$ complex absorption band is an OH stretch feature near 2.9 µm, which is not matched by the sloping Sedna band. Additionally, the broad 4- to 5-µm $H_2O$ feature is much weaker than on Gonggong and Quaoar. No Fresnel peak from crystalline $H_2O$ is apparent at 3.1 µm in Sedna's spectrum, but that location is dominated by the acetylene absorption feature described above. It is possible that crystalline $H_2O$ is present but masked by acetylene. Similarly, the region of 1.65-µm crystalline $H_2O$ band shows absorption from other materials that prohibit the assessment of the phase of $H_2O$. Future higher spectral resolution data may resolve individual absorption features in the 1.65-µm region.

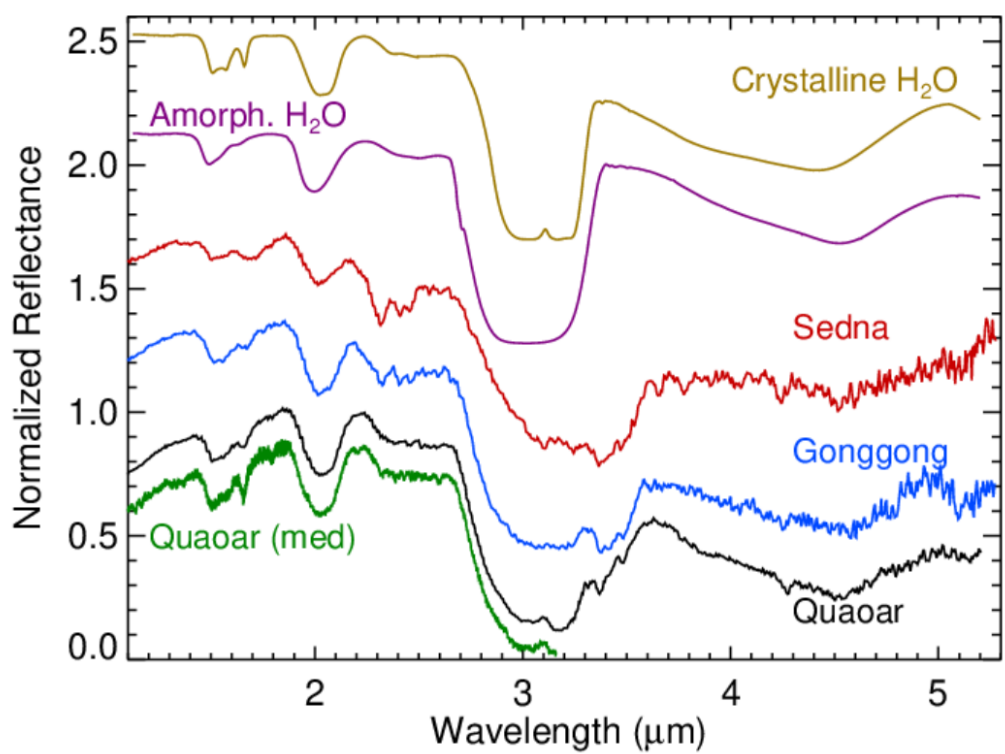

**Figure 5.** *The spectra of Sedna, Quaoar, and Gonggong compared to crystalline and amorphous $H_2O$. See the text for a discussion of the comparisons. The spectra of $H_2O$ are computed from optical constants (in a particulate mixture with 10% amorphous carbon) using Hapke theory. Optical constants are from Mastrapa et al. (2008, 2009).*

## 4.2 Gonggong

The JWST spectrum of Gonggong also shows strong evidence for abundant $C_2H_6$ on the surface. Several absorption features in the 3.35–3.5 μm range line up well with ethane features also present in the Sedna spectrum (Figs. 2, 3; Table 4). Additionally, the Gonggong spectrum contains a feature consistent with ethane at 3.42 μm that is not apparent in the Sedna spectrum, and the feature interpreted as due to ethane near 3.49 μm is at a slightly longer wavelength than in the Sedna spectrum. The slight difference in wavelength could be due to differences in mixing with other ices on the two bodies. Many hydrocarbons have absorptions at these wavelengths, and an alternative explanation could be that a different hydrocarbon ice is also present on Gonggong. For instance, propane ($C_3H_8$) has bands that match the Gonggong spectrum reasonably well in the 3.35–3.5 μm range. However, the absorptions in the 2.25–2.5 μm region on the two bodies do align very well with each other and match $C_2H_6$ features as described in Section 4.1 (Fig. 3). Many of the other features attributed to $C_2H_6$ in the Sedna spectrum, particularly those at $\lambda > 3.6$ μm, are absent from the Gonggong spectrum. The absence of these features, and weaker bands at 2.25 – 2.6 μm, suggest a lower relative abundance of $C_2H_6$ on Gonggong.

**Table 4.**
Band parameters and assignments for features identified in the spectrum of Gonggong. See text (Section 4) for details about calculations and uncertainties.

| $\lambda_c$ (μm) | | | FWHM (μm) | | | Band Depth (% of local cont.) | | | Assignment |
|---|---|---|---|---|---|---|---|---|---|
| ~1.5 | | | | | | | | | $H_2O$ |
| 1.674 | ± | 0.004 | 0.046 | ± | 0.011 | 4.0 | ± | 0.7 | Crystalline $H_2O$ |
| 2.023[a] | ± | 0.001 | 0.165 | ± | 0.003 | 27.5 | ± | 0.4 | $H_2O$ |
| 2.319 | ± | 0.001 | 0.054 | ± | 0.008 | 8.9 | ± | 0.6 | $C_2H_6$ |
| 2.408 | ± | 0.001 | 0.030 | ± | 0.005 | 6.3 | ± | 0.7 | $C_2H_6$ |
| 2.463 | ± | 0.002 | 0.023 | ± | 0.003 | 4.1 | ± | 0.6 | $C_2H_6$ |
| 2.594 | ± | 0.003 | 0.009 | ± | 0.019 | 3.6 | ± | 0.9 | ($C_2H_2$ or $CH_4$?) |
| 3.375[c] | ± | 0.002 | 0.032 | ± | 0.006 | 43.8 | ± | 4.9 | $C_2H_6$ |
| 3.419[c] | ± | 0.005 | 0.015 | ± | 0.027 | 26.5 | ± | 11.9 | $C_2H_6$ |
| 3.488[b,c] | ± | 0.003 | 0.030 | ± | 0.035 | 23.9 | ± | 10.3 | $C_2H_6$ |
| 4.253 | ± | 0.008 | 0.024 | ± | 0.033 | 24.0 | ± | 11.0 | Complexed $CO_2$ |

[a]These bands are asymmetric, and the Gaussian fits may not precisely represent the true center.
[b]This band is on a steeply sloped continuum, and the fit may not be precisely reliable.
[c]These bands are within the deep and broad 2.7 to 3.6-μm feature. The FWHM and Band Depth should be taken with consideration that only a local continuum was used.

There is very little evidence for $CH_4$ in the spectrum of Gonggong (see Fig. 3). No features are apparent in the 3.3- to 3.9-μm region that match with the strong $CH_4$ fundamental band at 3.32 μm or the relatively strong overtone and combination bands at 3.55 and 3.85 μm, and Gonggong does not show the small depressions that might be weak indications of $CH_4$ overtone and combination bands in the Sedna spectrum. One exception is the weak 2.60 μm feature that is also seen in the Sedna spectrum. Other $CH_4$ overtone and combination bands overlap with ethane features, and spectral modeling could reveal if small amounts of $CH_4$ could contribute to those absorptions. Gonggong also shows no evidence for the 3.1 μm band assigned to acetylene ($C_2H_2$) in the Sedna spectrum or the 3.24 μm feature assigned to ethylene ($C_2H_4$). In addition to the light hydrocarbon ices, complex organic molecules are indicated by the steep red spectral slope and also likely contribute to the broad absorption complex between 2.7 and 3.6 μm.

Gonggong has a feature at 4.253 μm close to the region of the $CO_2$ fundamental absorption. This feature is at a noticeably longer wavelength than the 4.243-μm feature in Sedna's spectrum and the corresponding $C_2H_6$ band suggested to be responsible for Sedna's feature (Fig. 4). It is also significantly weaker than the feature in Sedna's spectrum. Given the poor wavelength fit to $C_2H_6$ and that Gonggong does not show the other $C_2H_6$ bands in this spectral region, some of which are stronger than the ~4.24-μm band, it is unlikely that this feature is due to $C_2H_6$ in Gonggong's spectrum. Nevertheless, the feature is at a shorter wavelength than the $CO_2$ ice band (Fig. 4), and so may be due to $CO_2$ complexed in some way with the dark material or mixed with other ices as seen on Ganymede, Callisto and some of Saturn's moons (e.g., Hibbitts et al. 2000, 2003, Cruikshank et al. 2010).

The 1.5- and 2.0-μm features in the Gonggong spectrum look more like traditional $H_2O$ ice bands (Fig. 5) than those in the spectrum of Sedna. The 1.65-μm feature indicates the presence of the crystalline phase of $H_2O$ ice, which is also consistent with the position and symmetry of the 2.0-μm feature. The broad 2.7–3.6 μm absorption also better matches the shape of $H_2O$ ice than does the Sedna spectrum. The broad dip in the Gonggong spectrum in the ~4- to 5-μm region is also well matched to the shape of the $H_2O$ combination ice band. If the $H_2O$ ice is dominated by the crystalline phase, as indicated by the presence of the 1.65-μm feature, it is curious that the Fresnel peak at 3.1 μm is absent. One possible explanation is that the $H_2O$ right at the surface is dominated by amorphous rather than crystalline ice (i.e., layering occurs, as on Europa; Hansen and McCord 2004), or that the topmost layer is devoid of $H_2O$. Another possibility may be that $C_2H_2$ is present and its 3.1 μm band perfectly overprints the Fresnel peak. Detailed spectral modeling may distinguish those possibilities.

### 4.3 Quaoar

The JWST spectrum of Quaoar is more similar to the spectrum of Gonggong than Sedna, but there are also important differences from both that are discussed in this section. Additionally, as noted above, we measured a spectrum of Quaoar from 1 to 3.2 μm using the medium-resolution gratings (G140M and G235M), with resolving power from 4 to 25 times higher than the prism. Those data reveal additional weak absorptions in that range not resolved in the prism spectrum. The identification of these features raises the likelihood that additional diagnostic features could be resolved on Sedna and Gonggong if observed at higher spectral resolution. Table 5 gives the band parameters for Quaoar from both the prism and medium-resolution grating spectra.

**Table 5.**
Band parameters and assignments for features identified in the spectrum of Quaoar. See text (Section 4) for details about calculations and uncertainties.

| $\lambda_c$ (μm) | FWHM (μm) | Band Depth (% of local cont.) | Assignment |
|---|---|---|---|
| Prism mode | | | |
| ~1.5 | | | $H_2O$ |
| 1.661 ± 0.001 | 0.042 ± 0.003 | 4.9 ± 0.3 | Crystalline $H_2O$ |

| | | | | |
|---|---|---|---|---|
| 2.033 ± 0.001 | 0.182 ± 0.002 | 29.0 ± 0.2 | $H_2O$ |
| 3.103[b] ± 0.002 | 0.047 ± 0.005 | 25.7[b] ± 2.2 | $H_2O$ Fresnel (Crystalline) |
| 3.192[c] ± 0.001 | 0.136 ± 0.005 | 54.2 ± 1.5 | (HCN?) |
| 3.373[c] ± 0.002 | 0.033 ± 0.005 | 19.4 ± 1.8 | $C_2H_6$ |
| 3.487[c] ± 0.002 | 0.027 ± 0.005 | 10.1 ± 1.4 | $C_2H_6$ |
| 3.867 ± 0.004 | 0.165 ± 0.010 | 8.2 ± 0.4 | ? |
| 4.268 ± 0.002 | 0.044 ± 0.005 | 15.5 ± 1.1 | $CO_2$ ice |
| 4.680 ± 0.004 | 0.018 ± 0.027 | 10.6 ± 2.2 | CO? |
| **G140M/F100LP** | | | |
| 1.0921 ± 0.0015 | 0.0199 ± 0.0036 | 1.8 ± 0.3 | ? |
| 1.1364 ± 0.0005 | 0.0126 ± 0.0013 | 3.7 ± 0.3 | $CH_4$? |
| 1.1652 ± 0.0005 | 0.0131 ± 0.0013 | 3.2 ± 0.3 | $CH_4$? |
| 1.3077[a] ± 0.0020 | 0.0297 ± 0.0282 | 2.3 ± 2.2 | ? |
| 1.3341 ± 0.0006 | 0.0044 ± 0.0104 | 2.5 ± 0.3 | $CH_4$? |
| 1.3674 ± 0.0009 | 0.0169 ± 0.0033 | 2.4 ± 0.2 | $CH_4$? |
| ~1.5 | | | $H_2O$ |
| 1.6576 ± 0.0003 | 0.0286 ± 0.0010 | 10.9 ± 0.3 | Crystalline $H_2O$ |
| 1.7953 ± 0.0010 | 0.0170 ± 0.0018 | 3.2 ± 0.3 | $CH_4$? |
| **G235M/F170LP** | | | |
| 1.7885 ± 0.0012 | 0.0311 ± 0.0024 | 2.5 ± 0.2 | $CH_4$? |
| 2.0284 ± 0.0004 | 0.1817 ± 0.0008 | 30.0 ± 0.1 | $H_2O$ |
| 2.2076 ± 0.0015 | 0.0162 ± 0.0040 | 1.9 ± 0.3 | $CH_4$ |
| 2.3198 ± 0.0007 | 0.0264 ± 0.0035 | 4.0 ± 0.3 | $C_2H_6$ and $CH_4$ |
| 2.3848 ± 0.0007 | 0.0111 ± 0.0025 | 3.1 ± 0.4 | $CH_4$ |
| 2.4049 ± 0.0006 | 0.0019 ± 0.0048 | 2.3 ± 1.5 | $C_2H_6$ |
| 2.4292 ± 0.0005 | 0.0103 ± 0.0015 | 3.3 ± 0.4 | $C_2H_6$ |
| 2.4607 ± 0.0004 | 0.0069 ± 0.0008 | 3.7 ± 0.4 | $C_2H_6$ |
| 2.5983 ± 0.0010 | 0.0188 ± 0.0025 | 3.3 ± 0.4 | $CH_4$ |
| 3.0986 | | | $H_2O$ Fresnel (Crystalline) |
| [b] ± 0.0018 | 0.0437 ± 0.0041 | 17.2[b] ± 1.5 | |

[a]These features are asymmetric, and the Gaussian fits may not precisely represent the true center.
[b]This feature is a peak (the crystalline $H_2O$ Fresnel peak). Depth is a height above continuum.
[c]These bands are within the deep and broad 2.7- to 3.6-µm feature. The FWHM and Band Depth should be taken with consideration that only a local continuum was used.

The fundamental C-H stretching features at ~3.37 and 3.49 µm are weaker in the spectrum of Quaoar than in the spectra of Sedna and Gonggong (Fig. 2). Nevertheless, the absorption features are at the same wavelengths as in the Sedna spectrum and are consistent with $C_2H_6$. The additional longer-wavelength features of $C_2H_6$ seen in the spectrum of Sedna are not apparent in the spectrum of Quaoar, which is consistent with the lower abundance of $C_2H_6$ suggested by the weaker fundamental bands at ~3.4 and 3.5 µm. At shorter wavelengths, the set of absorptions in the ~2.25- to 2.6-µm region are far weaker in the Quaoar spectrum than in the spectra of the other two objects. In fact, they are so weak that they are not apparent in the prism spectrum and are only seen in the medium grating spectrum (Fig. 6a). These higher spectral resolution data reveal weak features at ~2.320, 2.405, 2.429, and 2.461 µm that are consistent with $C_2H_6$. Schaller and Brown (2007a) and Dalle Ore et al. (2009) identified features at ~2.275 and 2.315 µm in ground-based spectra with $C_2H_6$. Their ~2.275-µm feature is not readily apparent in the JWST spectrum, but may be present as a slight depression on the side of the continuum defined by the 2-µm $H_2O$ band. The 2.429-µm feature is at a slightly longer wavelength than the corresponding ethane band (2.424 μm) and is stronger relative to surrounding bands than expected for ethane, so there may be another material contributing here as well. The band centers of the $C_2H_6$ features detected on Quaoar are consistent with pure ethane, showing no indications of shifts that would suggest dissolution in $N_2$.

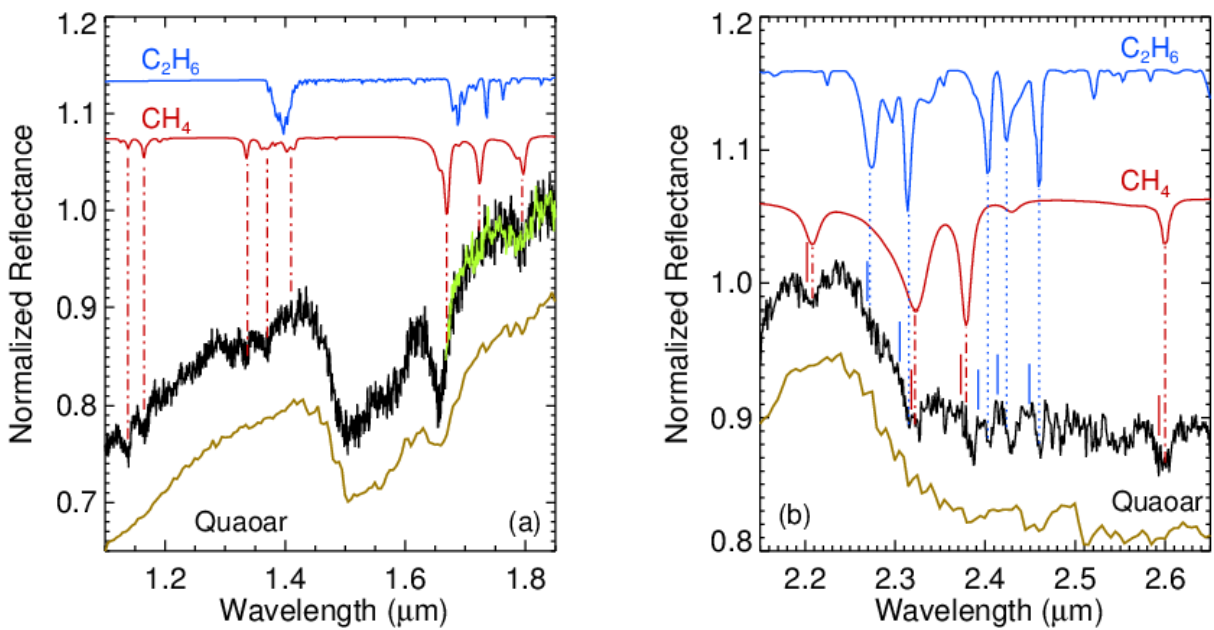

**Figure 6.** *Close-up view of the (a) 1.1-1.85 μm region and (b) 2.15-2.65 μm region of the medium resolution grating spectrum of Quaoar compared with spectra of ethane and methane ices. Features in the spectrum line up with both ethane and methane bands in the 2.15-2.65 μm region (b). In the shorter wavelength range (a), some features appear to align with some methane bands, but other methane bands are absent. The bright green line overplotting the spectrum in (a) from 1.66 to 1.85 μm is the overlapping segment of the G235M/F140LP part of the spectrum. The dark yellow curve in both frames is the lower resolution prism spectrum, for comparison. Note that both orders show the 1.795 μm feature. The short vertical red and blue lines in (b) show the positions of $CH_4$ and $C_2H_6$, respectively, dissolved in $N_2$ (from Quirico and Schmitt 1997). Shifts are smaller in the 1.1-1.85 μm region and are not shown. The methane and ethane bands observed on Quaoar more closely match the pure species, showing no indication of being dissolved in $N_2$.*

As with Sedna and Gonggong, the $CH_4$ fundamental band at 3.32 μm is not apparent in the spectrum of Quaoar, nor are the overtone and combination bands at 3.55 and 3.85 μm. Determination of whether or not a small amount of $CH_4$ could contribute to the overall 2.7- to 3.6-μm absorption will require spectral modeling. In the 2.2- to 2.6-μm region, the medium grating spectrum of Quaoar shows four absorptions consistent with methane: ~2.208, 2.320, 2.385, and 2.600 μm (Fig. 6a), though the 2.385-μm feature in the Quaoar spectrum is at a slightly longer wavelength than the corresponding methane band (2.379 μm). The band centers of these $CH_4$ features detected on Quaoar are consistent with pure methane, particularly the 2.208- and 2.600-μm features, showing no indications of shifts that would suggest that the $CH_4$ is dissolved in $N_2$. Furthermore, absorptions at 1.136, 1.165, 1.334, 1.367, and 1.795 μm are also consistent with CH$_4$ (Fig. 6b), and each of these bands are labeled as *strong* relative to surrounding bands by Calvani et al. (1992). However, not all methane bands in this wavelength range are apparent in the Quaoar spectrum, and the two shortest wavelength features appear stronger than expected for methane relative to the other features. Schaller and Brown (2007a) reported detection of the ~2.208-, 2.320-, and 2.385-µm methane bands, and Jewitt and Luu (2004), Dalle Ore et al. (2009), and Barucci et al. (2015) also discuss detection of the ~2.208-µm band. All previous authors note that ammonia hydrate also has a band near 2.21 µm, but also discuss that the presence of the other methane bands argues for the assignment to methane rather than ammonia hydrate. Detailed spectral modeling could test whether there could be any contribution from ammonia hydrate.

Not all of the features in the medium grating spectrum are perfectly matched in shape, width, relative depth, and/or position by methane or ethane, suggesting that a closer analysis may reveal contributions from other hydrocarbons as well. The presence of CH$_4$ spectral features at short wavelength together with the absence of the stronger absorptions at longer wavelengths may indicate compositional layering in Quaoar's near surface. Because the absorption coefficients of detected surface ices are smaller at shorter wavelengths, photons would be expected to penetrate deeper into the surface at shorter wavelengths ($\lambda < 2.7$ µm) than at longer wavelengths. In addition to the light hydrocarbon ices, complex organic molecules are indicated by the steep red spectral slope and also likely contribute to the broad absorption complex between 2.7 and 3.6 µm.

Quaoar shows a strong absorption feature at ~3.19 µm that is not present in the spectra of Sedna or Gonggong. Though the 3.1-µm Fresnel peak indicative of crystalline H$_2$O ice is present, the shape at 3.2 µm is not consistent with H$_2$O. The location of this 3.19-µm absorption is reasonably well matched to HCN. However, HCN also has a strong band at ~4.76 µm that does not appear in the spectrum of Quaoar (Fig. 7). Perhaps the 4.76-µm feature is somehow masked in the mixing of materials on the surface. We also considered H$_2$O to explain this 3.19-µm feature. The 3.1-µm Fresnel peak indicative of crystalline H$_2$O ice is present in the spectrum of Quaoar, and the broad 3-µm band includes lobes at both shorter and longer wavelengths than the Fresnel peak. However, in the spectrum of H$_2$O ice, the longer wavelength lobe (which is centered at ~3.19 µm) is generally equally (Fig. 5) or *less* (e.g., Clark et al. 2008, 2012) absorptive than the shorter wavelength lobe. We have tried modeling very small H$_2$O grains (using Mie theory to compute the single scattering albedo, which was then inserted into the Hapke reflectance equation), thin frost coatings on grains (e.g., Rivkin and Emery 2010), and intraparticle mixtures, but none of these scattering schemes produced a deeper 3.19-µm lobe in the H$_2$O feature. We also compared Quaoar's 3.19-µm feature with absorptions due to ammoniated species detected on the Main Belt dwarf planet Ceres and the Jupiter Family Comet 67P/Churyumov-Gerasimenko (e.g., Poch et al. 2020). The ammoniated clays detected on Ceres have a band center near 3.07 µm (e.g., De Sanctis et al. 2018), and the feature due to ammonium salts detected on 67P is centered at about 3.1 µm (e.g., Poch et al. 2020). Neither feature provides a match to the 3.19-µm absorption in Quaoar. Additionally, Cerealia Facula (the bright dome within Occator Crater) shows evidence for chloride salts, but those show a double-lobed band with an absorption maximum near 3.29 µm and an additional absorption at about 3.52 µm (De Sanctis et al. 2020) not seen in the Quaoar spectrum. Poch et al. (2020) show a spectrum of NH$_4$Cl that has a peak absorption at 3.19 µm, but it also shows a strong second absorption at about 3.3 µm that is not seen in Quaoar's spectrum. Nevertheless, it would be useful to explore other salts as possible candidates. For now, we consider HCN as the best explanation of the 3.19-μm feature in Quaoar's spectrum, though the absence of the 4.76-μm band is problematic.

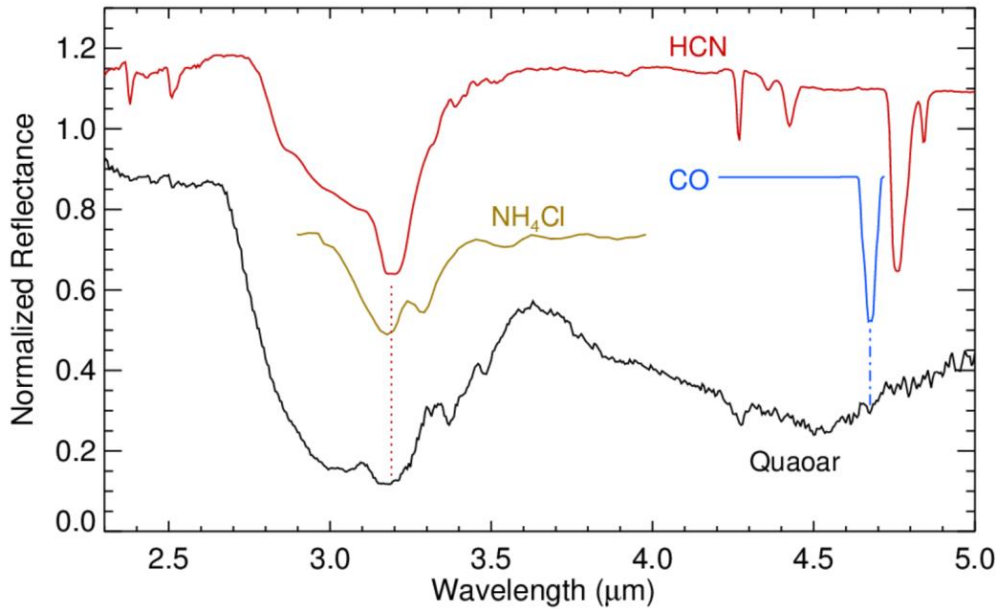

***Figure 7.*** *Comparison of the prism spectrum of Quaoar with spectra of HCN and, CO and NH₄Cl. The ~3.19-μm feature in the spectrum of Quaoar is possibly matched by the corresponding band in HCN, but HCN has a strong band near 4.77 μm that is not seen in the spectrum of Quaoar. NH₄Cl also has a band at approximately the right position, but a second feature near 3.29 μm is absent from the spectrum of Quaoar. A possible feature at 4.675 μm in the Quaoar spectrum is in the right place for CO ice, but the feature is at the level of noise in the data. The spectra of HCN and CO are computed from optical constants (in particulate mixtures with 10% amorphous carbon and grain sizes 2.5 and 7.5 μm, respectively) using Hapke theory. Sources of optical constants are: HCN - Moore et al. (2010); CO - R.H. Brown (personal communication). The spectrum of NH₄Cl is from Poch et al. (2020).*

A feature at 4.268 μm in the spectrum of Quaoar is at a slightly longer wavelength than the similar features noted above for Sedna and Gonggong. In the case of Quaoar, this feature is in the right location for the $CO_2$ ice fundamental band (Fig. 4). The absorption coefficients for $CO_2$ ice are very large in this band, and the appearance of the feature as a relatively weak, more-or-less Gaussian shaped band suggests very fine grained $CO_2$ ice frost in low relative abundance. The weaker $CO_2$ bands at ~2.7 and 2.8 μm (which are seen in some other KBO spectra; de Prá et al. 2024, Brown and Fraser 2023, Pinilla-Alonso et al. 2024) are not visible in the Quaoar spectrum, consistent with the relatively weak fundamental band. A possible feature near 4.680 μm is in the right location for CO ice, but the noise in the spectrum at that wavelength makes the identification uncertain (Fig 7).

The shapes of the 1.5- and 2.0-μm features in the spectrum of Quaoar are consistent with relatively pure $H_2O$ (Fig 5). As with Gonggong, the 1.65-μm feature is indicative of crystalline water ice. In the prism spectrum, this feature is weaker relative to the rest of the 1.5-μm feature than observed in ground-based spectra (Jewitt and Luu 2004, Dalle Ore 2009, Barucci 2015). In the medium grating spectrum, however, the relative depths are similar to the ground-based data. By analogy, the 1.65-μm features for Gonggong and Sedna may be deeper than indicated by the low-resolution prism spectra. The spectral resolution (Δλ) of prism mode is only Δλ~0.02 μm at this wavelength, so there are only one or two resolution elements on the 1.65-μm band in the prism spectra. A peak in the Quaoar spectrum at ~3.1 μm may be the Fresnel peak from crystalline $H_2O$. The broad bowl-shape from ~3.7 to 5.0 μm is also consistent with $H_2O$.

The Quaoar spectrum contains a somewhat shallow (~7%) dip relative to the local continuum at ~3.75 to 3.95 μm (Fig. 8). $H_2S$ has an absorption band close to this wavelength, but the $H_2S$ band is centered slightly longward of the feature in the Quaoar spectrum. Garozzo et al. (2010) show the $H_2S$ band is shifted shortward to 3.89 μm when mixed as a minor component with CO (CO:$H_2S$=10:1), and Moore et al. (2007) show a similar shift to 3.90 μm when mixed with $H_2O$ ($H_2O$:$H_2S$=11:1). Those shifts are not sufficient to explain the position of the feature seen in the spectrum of Quaoar, but perhaps mixing with ethane and/or methane could produce a greater shift. Carbonates also have absorption bands near this wavelength, but we have not identified a carbonate with a band of the right location and shape. The peak absorption in the carbonate spectra we looked at (from the ASTER[2] and USGS[3] databases) is always longward of 3.9 μm.

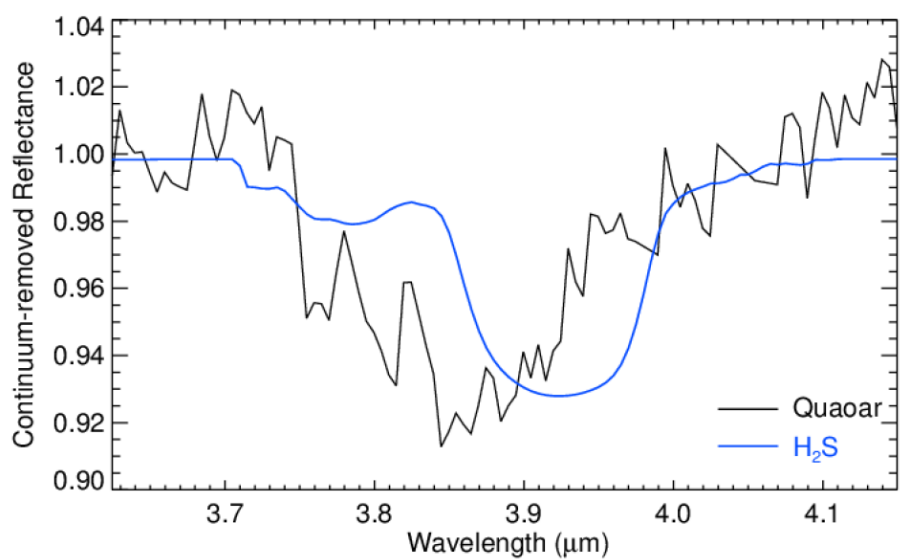

**Figure 8.** *Close-up of the Quaoar JWST NIRSPEC prism spectrum showing a shallow and broad absorption near 3.85 μm as compared to the spectrum of $H_2S$. A local continuum was defined by a linear fit between 3.62-3.74 μm and 3.95-4.15μm, and the Quaoar spectrum was divided by that continuum. The feature is not quite at the right position for $H_2S$. The spectrum of $H_2S$ is computed from optical constants (in a particulate mixture with 10% amorphous carbon and 5-μm grain size) using Hapke theory. Optical constants from Trotta and Schmitt (1992).*

---

*4.4 Notable species not detected*

Several species hypothesized or suggested to be present on KBO surfaces are not detected in the JWST spectra of these three dwarf planets. Methanol (CH₃OH) has been suggested to be widespread in the outer Solar System, based on its presence in comets and on a ~2.27-μm band detected in spectra of the centaur (5145) Pholus (Cruikshank et al. 1998) and the KBOs (55638) 2002 VE95 (Barucci et al. 2006) and 486598 Arrokoth (Grundy et al. 2019). None of the spectra of the three objects presented here show evidence for any methanol features (Fig. 9).

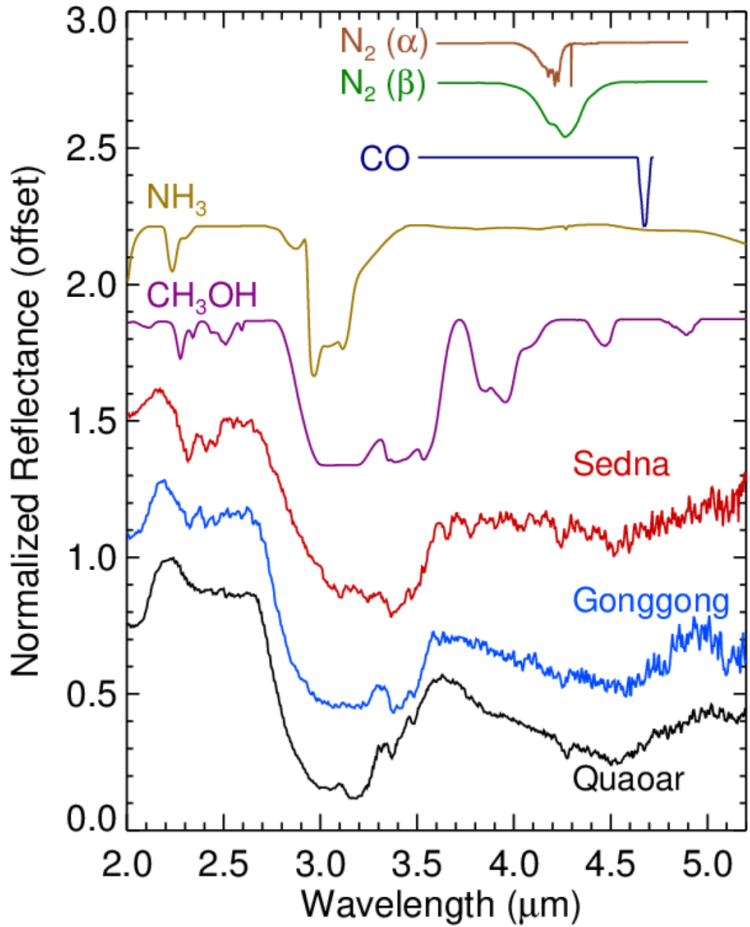

**Figure 9.** *Comparison of the spectra of Sedna, Gonggong, and Quaoar with those of CH₃OH, NH₃, N₂, and CO ices. Their spectral features are not detected in the JWST spectra presented here. The spectra of CH₃OH, NH₃, N₂, and CO are computed from optical constants (in particulate mixtures with 20, 10, 10, and 10% amorphous carbon and grain sizes 7.5, 2.5, 500, and 7.5 μm, respectively) using Hapke theory. Sources of optical constants are: NH₃ - Hudson et al. (2022); CH₃OH and CO - R.H. Brown (personal communication); N₂ (α) - Schmitt et al. (1998); N₂ (β) - Green et al. (1991).*

N₂ might reasonably be expected on these bodies, particularly at the very cold temperatures on Sedna, given its presence on the largest dwarf planets. Previous modeling of spectra of Sedna and Quaoar have argued for inclusion of N₂ to improve fits to the 3.6- and 4.5-μm *Spitzer* IRAC broadband photometric measurements (Emery et al. 2007, Barucci et al. 2008,

Dalle Ore et al. 2009, Barucci et al. 2015), and Barucci et al. (2005) presented a possible absorption from $N_2$ on Sedna at 2.15 μm (though the spectral region was noisy and additional ground-based data presented in Barucci et al. (2010) do not show that absorption). However, we see no evidence for either the ~2.15-μm or ~4.25-μm $N_2$ absorptions in any of the spectra presented here. $N_2$ undergoes a phase transition at 35.6 K, near its boiling point at 1 Pa (Fray and Schmitt (2009), which somewhat changes the shapes of the absorption bands (Fig. 9). For the warmer temperature β phase, the 4.25-μm band is much too broad to explain the feature located near that wavelength in the measured spectra. The lower temperature α phase resolves into a very sharp, vibrational band at 4.296 μm and a broader lobe of multi-phonon bands from about 4.15 to 4.24 μm (Quirico et al. 1996, Brodyanski et al. 2002). The vibrational band is too narrow to be seen in the JWST prism data, and the multi-phonon bands are too broad, extending to shorter wavelengths than the observed feature in the measured KBO spectra. The methane and ethane features we detect in the medium grating spectra of Quaoar are at positions consistent with their occurrence as pure ices on the surface. There is no indication of band shifts that would be indicative of dissolution of these species in $N_2$. Barucci et al. (2015) argue that the position of the 1.65-μm crystalline $H_2O$ band in their ground-based spectrum is affected by methane in a way that suggests a shifted methane band (with a comparison to such an effect in the spectrum of Triton). However, the strengths of the methane bands on Quaoar at these wavelengths are too weak to affect the much stronger $H_2O$ band. The situation is reversed on Triton, where the methane absorptions dominate over the water ice absorptions in the spectrum. The spectral resolution of the prism spectra is too low to assess whether or not the ethane bands detected on Sedna and Gonggong are shifted.

NH$_3$ or related materials have been detected on icy moons (e.g., Cartwright et al. 2020, 2023) and Charon (Brown and Calvin 2000, Cook et al. 2022, Protopapa et al. 2023), and have been suggested in spectra of other KBOs (e.g., Barucci et al. 2008, Dalle Ore et al. 2009) based, in part, on improved fits of spectral models including such materials. We see no specific features attributed to NH$_3$-related materials in the spectra of Sedna, Gonggong, or Quaoar (Fig. 9). Future spectral modeling will investigate whether NH$_3$-related hydrates may help explain the spectral shape around the 2.2-μm peak, including the absorption at 2.208 μm in Quaoar's spectrum that we attributed to CH$_4$.

We mention in section 4.3 a possible feature at the location of the CO fundamental stretching band in the spectrum of Quaoar (Fig. 7, Table 5), but the detection is tentative, and we see no similar feature in the spectra of Sedna or Gonggong. CO is detected in JWST spectra of smaller KBOs that also show very strong $CO_2$ absorptions (Pinilla-Alonso et al. 2024, Brown and Fraser 2023). Brown and Fraser (2023) suggest the CO is produced by irradiation of $CO_2$, then trapped within the $CO_2$, which provides stability. If that scenario is correct, the absence of CO on the surfaces of Sedna and Gonggong could be related to the absence of $CO_2$ ice at the surface. Birch and Umurhan (2023) recently suggested a scenario where primordial CO may be stabilized in the interiors of small, cold KBOs by vapor pressure in overlying porespace. CO is also absent from spectra of Eris and Makemake and is orders of magnitude less abundant on Pluto than expected from comets. This "missing CO" can be explained by low original abundances of supervolatiles (with the presently observed $N_2$ representing a secondary reservoir), hydrothermal destruction of CO in a warm interior, or some form of sequestration below the surface (McKinnon et al. 2021). In any case, the low abundance of CO on Sedna, Gonggong, and Quaoar may suggest that they are geochemically comparable to Pluto, Eris, and Makemake rather than the smaller KBOs.

## 5. Discussion

### 5.1 Ethane abundance and volatile retention

Based on the number of detected ethane bands in each spectrum and their relative depths, ethane appears to be most abundant on Sedna and least abundant on Quaoar, with Gonggong in between. The ethane on these bodies most likely formed by UV and particle irradiation of methane via dehydrogenation and radical cross-linking (Gerakines et al 2001, Bennett et al. 2006) Upon continued irradiation, by either UV photons or high energy charged particles, these materials evolve to more carbon rich (as H is lost), long chain organic molecules (e.g., Thompson et al. 1987, Moroz et al. 2004, Bennett et al. 2006, Quirico et al. 2023, Zhang et al. 2023). The inference that ethane formed by irradiation of methane is supported by the detection of acetylene and ethylene, additional products of $CH_4$ irradiation, in the spectrum of Sedna. $C_2H_6$, $C_2H_4$ and $C_2H_2$ are all expected to form by radiolysis of methane, in decreasing abundance (Bennet et al. 2006, about 3 to 5 times less $C_2H_4$ than $C_2H_6$, and about 10 to 20 times less $C_2H_2$ than $C_2H_6$). The inference that $C_2H_6$ is formed by irradiation is also supported by the absence of $C_2H_6$ in the inventory of ices recently reported from JWST observations of very dense inner regions of the molecular cloud around the young star-forming core Cha MMS1 (McClure et al. 2023). The question then arises whether the observed differences of ethane abundances on these bodies are due to different original complements of $CH_4$ or different abilities of each to retain $CH_4$ for the production of $C_2H_6$ by irradiation.

The observed $C_2H_6$ abundances correlate with orbital properties, most notably semi-major axis and aphelion distance (Table 1). Stability of volatile ices is also dependent on heliocentric distance, and it is useful to consider the stabilities of species on these three bodies. $C_2H_6$ has a relatively low volatility and is stable on all three bodies (e.g., Brown et al. 2011b), so the trend in ethane abundance must be related to its formation. Acetylene and ethylene have roughly similar volatilities as ethane, so would also be retained on these bodies once formed. The stability of methane, however, varies among these three bodies. Models of Jeans escape from KBOs predict that $CH_4$ should be stable on Sedna, whereas Gonggong and Quaoar straddle the stability line (Schaller and Brown 2007b, Brown et al. 2011a, Zahnle and Catling 2017). Quaoar has a fairly circular orbit with the smallest semi-major axis of the three, so could be expected to undergo relatively steady low-level loss of $CH_4$. For Gonggong, because of the highly non-linear dependence of sublimative loss with heliocentric distance, the vast majority of its $CH_4$ loss would occur rapidly near perihelion. Johnson et al. (2015) model both surface and atmospheric heating for loss of $N_2$ from these bodies and find similar loss rate trends.

The longer the methane stays around to be irradiated, the more ethane (and acetylene and ethylene) should be produced. The trend of ethane abundance is therefore consistent with long-term irradiation of methane, where the better retention of $CH_4$ determines abundance of ethane among these three bodies. In other words, the differences in ethane abundances in these three bodies appear to be evolutionary, not indicative of different primary abundances of methane. In addition, the larger the aphelion, the longer the residence in regions of the Solar System where ion fluxes are very high (in particular beyond the termination shock at ~85 AU and the heliopause at ~120 AU, see Strazzulla et al. 2003, Cooper et al. 2003). This is particularly true for Sedna, which can easily accumulate the few eV to tens of eV per molecule that are necessary to produce $C_2H_6$, $C_2H_4$, and $C_2H_2$ at its surface starting from $CH_4$ (Bennet et al. 2006 and references therein, see discussion below).

The source of the methane itself on these bodies is also an interesting question. Grundy et al. (2023) report D/H and $^{13}C/^{12}C$ ratios on the dwarf planets Eris and Makemake from recent JWST spectral observations. The measured D/H is similar to carbonaceous chondrite meteorites and other large outer Solar System bodies and much lower than that of presumably primordial comet reservoirs. They argue that these values indicate that the methane on Eris and Makemake is not primordial, but results from internal geochemical processing, probably in the presence of liquid water. Gonggong and Quaoar have similar densities to Makemake, Eris, Pluto, and Triton, suggesting similar rock mass fractions (e.g., Bierson and Nimmo 2019). No moon has been detected around Sedna, so it is not yet possible to calculate its density, but with a similar size to Gonggong, it is reasonable to assume that it may also have a similar density and rock mass fraction. It is plausible that Sedna, Gonggong, and Quaoar are also large enough to have undergone internal heating, differentiation, and associated geochemical processing. The $CH_4$ reservoir that was irradiated to form ethane and other hydrocarbons on Sedna, Gonggong, and Quaoar may, therefore, have also been sourced geochemically.

One striking difference between the three dwarf planets in our study and Pluto, Eris, Triton, and Makemake is the abundance of methane spectrally detected. The latter four bodies are all large enough to gravitationally retain $CH_4$. Furthermore, atmospheric and geologic processes likely refresh their icy surfaces, keeping their albedos relatively high (e.g., Buratti et al. 2017, Verbiscer et al. 2022). Higher methane loss rates on Quaoar and Gonggong, combined with less atmospheric and geologic activity refreshing their and Senda's surfaces may conspire to account for the smaller methane abundances and higher ethane (and other hydrocarbon) abundances than their larger kin. Another distinction is the detection of $N_2$ ice on Pluto, Triton, and Eris, but not on the smaller bodies. Johnson et al. (2015) find that $N_2$ on Quaoar would be depleted by ~88% over the lifetime of the Solar System, but only 18-31% would be lost from Gonggong and 1.8% from Sedna. $N_2$ may be produced from high temperature, oxidizing interior hydrothermal reactions (e.g., Glein 2023), but may not be produced at the lower temperatures achieved in the interiors of Makemake, Sedna, Gonggong, and Quaoar (e.g., Glein et al. 2008). Another possibility is that $N_2$ was formed on these bodies but is covered by irradiation products, but $N_2$ is very mobile, so we expect it more likely escaped if it was initially present on the surface. If $N_2$ were abundant as $CH_4$ is irradiated, we would expect to see N-related irradiation products, along with the light hydrocarbons identified in the JWST spectra. The absence of any clear N-related features provides another indication that $N_2$ is not abundant on these surfaces. However, the distorted $H_2O$ features in Sedna's spectrum require an explanation, and future spectral modeling of these features will address whether nitrogen bonds within the complex organics contribute.

The spectra of Sedna, Gonggong, and Quaoar are also distinct from those of smaller KBOs. Pinilla-Alonso et al. (2024) find that JWST NIRSpec/prism spectra of 54 KBOs cluster into 3 distinct spectral and compositional groups: a "bowl" group spectrally dominated by $H_2O$; a "double-dip" group that shows a deep, rounded 3-μm band and very strong $CO_2$ and and CO features; and a "cliff" group that shows a steeper drop into a wider 3-μm band with a flatter bottom, suggesting chemically complex organics. The three objects in our study are distinguished from those groups in the shapes of the overall 3-μm band and the relative weakness of the $CO_2$ bands. None of the smaller KBOs in Pinilla-Alonso et al. (2024) show prevalence of ethane in the spectra of the three objects reported here (nor the acetylene and ethylene on Sedna). By contrast, many of the "cliff" objects show features indicative of methanol, which is absent from the spectra presented in this paper (see section 4.4). The objects observed by Pinilla-Alonso et al. (2024) range in diameter from ~100 to ~800 km, and they show no size dependence in the spectral/compositional groupings. These differences of Sedna, Gonggong, and Quaoar from the smaller KBO population supports the inference of different evolutionary history based on their size.

*5.2 Other irradiation effects*

Sedna, Gonggong, and Quaoar all exhibit steep red spectral slopes at visible and NIR wavelengths. These steep spectral slopes are likely due to the presence of complex organics resulting from the irradiation of simple ices ("tholins", e.g., Cruikshank et al. 1998). Complex organics also likely contribute strongly to the broad and deep absorption from ~2.7 to 3.6 μm. Various organic residues have been produced from various starting ices and with different energy sources (e.g., Khare et al. 1984, McDonald et al. 1996, Brunetto et al. 2006, Javanović et al. 2021, Zhang et al. 2023), and they tend to share these spectral properties. Because they are composed of complex mixtures of complex molecules, their spectra tend to lack discrete absorptions such as those we have identified with simple hydrocarbons above. We therefore expect these surfaces to comprise a large suite of organic molecules created by irradiation of their surfaces, not just the relatively simple products discovered by feature identification. Organic residues can also exhibit absorption at 1.5 and 2.0 that are differently shaped than $H_2O$ bands, which may help explain the distorted appearance relative to pure $H_2O$ of the features at these wavelengths in Sedna's spectrum (e.g., Fayolle et al. 2021). The absence of $NH_3$ in the spectra of the three objects may also be explained by an efficient radiolysis of ammonia, which in turn makes the nitrogen atoms available to be incorporated together with carbon and hydrogen (and possibly some oxygen) into more complex tholin-like materials.

The different orbits of these three dwarf planets lead to different irradiation environments. With a perihelion distance of 76 AU and aphelion distance of 937 AU, Sedna spends most of its time outside of the heliopause (at ~120 AU; e.g., Stone et al. 2013, 2019), where the flux of galactic cosmic rays is significantly higher than inside the heliosphere. Resulting times to deposit chemically relevant doses (a few to a few hundred Myr, depending on depth) are estimated to be one to a few orders of magnitude shorter (Strazzulla et al. 2003, Cooper et al. 2003) than at 40 AU. Gonggong's aphelion of 101 AU takes it beyond the termination shock at ~85 AU (e.g., Stone et al. 2005, 2008), but not beyond the heliopause. Cosmic ray fluxes are significantly elevated in this region (called the heliosheath) as well, though not quite as much as beyond the heliopause (Strazzulla et al. 2003, Cooper et al. 2003), resulting in irradiation timescales of 100 Myr to 1 Gry, depending on depth. With a nearly circular orbit around 43.7 AU, Quaoar's orbit keeps it firmly within the more typical irradiation environment of the Kuiper Belt.

The observation of strong features from the immediate irradiation products of methane (i.e., ethane, acetylene, ethylene) suggests some resupply of methane to these surfaces. Furthermore, Sedna's relatively high visible-wavelength geometric albedo of 0.32 (which suggests a NIR geometric albedo close to or exceeding 1.0) indicates an ice-rich surface, despite the clear spectral signatures of complex organic molecules. Occultations of Quaoar detect no sign of an atmosphere, placing an upper limit of 10 to 20 nbar (Braga-Ribas et al. 2013, Arimatsu et al. 2019). Johnson et al. (2015) suggest Gonggong could develop a temporary atmosphere as its orbit brings it within ~70 AU of the Sun. Sedna is not expected to support an atmosphere at its great distances. Surface refreshing mechanisms are therefore likely different than on Pluto and Triton, but could include seasonal volatile migration, outgassing from the interior, or possibly some geologic activity.

## 5.3 Carbon Dioxide

As noted above, the absorption in the spectrum of Quaoar at 4.268 μm is consistent with $CO_2$ ice. In the spectra of Gonggong and Sedna however, the feature is shifted shortward. The absorption feature is centered at 4.253 μm for Gonggong. Similar shifts in the fundamental $CO_2$ band have been detected on Ganymede and Callisto (4.257 and 4.258 μm, respectively; Hibbitts et al. 2000, 2003) and on several satellites of Saturn (e.g., 4.252 and 4.253 μm on Dione and Hyperion, respectively; Cruikshank et al. 2010). Cruikshank et al. (2010) explain the wavelength shift as caused by the $CO_2$ molecules being complexed with other materials, such that $CO_2$ molecules are isolated from one another. They describe cases where the other materials are other ices and cases where it is refractory material. In mixing with different ices, Sandford and Allamandola (1990) find the largest shift is to 4.263 μm, which is not enough to explain the feature detected on Gonggong (and that mixture is a 20:1 mixture of $CO:CO_2$, but we also do not see a CO absorption on Gonggong). On Ganymede, Callisto, and several of the Saturnian satellites, the shifted bands are associated with low-albedo material, perhaps suggesting that $CO_2$ is complexed with that non-ice material. Gonggong's spectrum indicates the presence of $H_2O$ ice, ethane ice, and complex organic material, which could all be candidates for the material that $CO_2$ is complexed with.

In the case of Sedna, the absorption feature is centered at 4.243 μm and is significantly wider than on Gonggong. As described in section 4.1, the location of the feature on Sedna is well matched by a $C_2H_6$ absorption at 4.242 μm, which is consistent with the detection of many other $C_2H_6$ bands in the spectrum of Sedna. The feature on Sedna, however, is deeper relative to surrounding ethane bands and wider than expected for ethane. It is therefore possible that this feature has a contribution from both $C_2H_6$ and $CO_2$. In fact, the feature is wide enough that there could even be a contribution from $CO_2$ ice as well as (or instead of ) complexed $CO_2$. Spectral modeling will be required to test this possibility and disentangle the signatures. Once produced, $CO_2$, whether ice or complexed with other materials, would be relatively stable on all three surfaces.

A number of smaller objects in the Kuiper Belt show extremely strong, and somewhat anomalous, $CO_2$ bands (de Prá et al. 2024, Brown and Fraser 2023). Those detections indicate that $CO_2$ is common on smaller KBOs and can be present in large abundances. If the large abundances of surface $CO_2$ on those bodies is due to delivery from the interior (Brown and Fraser 2023), we might expect the large bodies studied here to have significant abundances of $CO_2$ on their surfaces. Ground-based spectra of Pluto, Eris, and Makemake, on the other hand, show no evidence for $CO_2$, nor do *New Horizons* spectra of Pluto (though the spectral resolution of LEISA was too low to detect small amounts of $CO_2$). $CO_2$ is refractory at the surface temperatures of all these large KBOs, and $CO_2$ could be covered up by volatile ices and/or products of the active irradiation environments, or it could indicate that there is some other explanation for the large $CO_2$ abundances on the smaller bodies on which it is observed (Pinilla-Alonso et al. 2024).  The weak $CO_2$ ice feature detected on Quaoar and the complexed $CO_2$ detected on Gonggong (and possibly Sedna) may indicate buried $CO_2$.

## 6. Summary and Conclusions

JWST spectra of Sedna, Gonggong, and Quaoar show a host of absorption features revealing new details about their surface compositions.

- Sedna's spectrum contains features due to $H_2O$, ethane ($C_2H_6$), acetylene ($C_2H_2$), ethylene ($C_2H_4$), complex organics (i.e., tholins), and possibly $CO_2$.  The presence of more than a dozen features attributable to ethane indicates a significant abundance of ethane on its surface.
- Gonggong's spectrum shows deeper $H_2O$ features, several ethane features (though fewer and weaker than on Sedna), complex organics, and $CO_2$ that is complexed with other molecules on the surface.
- Quaoar's prism-mode spectrum shows the cleanest $H_2O$ features, including an apparent Fresnel peak at 3.1 μm, ethane indicated only by the very strong bands near 3.37 and 3.45 μm, complex organics, $CO_2$ ice, a tentative detection of CO, and a feature at 3.2 μm that may be due to HCN, though the absence of the ~4.7-μm HCN band makes the identification uncertain.  The ~10x higher spectral resolution spectrum of Quaoar measured with the medium-resolution gratings additionally reveals overtone and combination bands of both $CH_4$ and $C_2H_6$ that are not detected in prism mode.

The rich complement of light hydrocarbons and complex organics detected indicates robust irradiation chemistry active on all three bodies.  We infer that these three dwarf planets are large enough to retain methane long enough to allow irradiation products to form.  Differences among the three are likely due to their different orbits.  Quaoar's more circular, hot-classical Kuiper Belt orbit probably leads to a relatively steady loss of $CH_4$ and thereby less time for irradiation products to build up.  Gonggong's more eccentric orbit provides more stability for $CH_4$ and therefore longer time for irradiation products to form.  The extension of its aphelion beyond the termination shock also affects the spectrum of energetic particles irradiating the surface.  Sedna's even more extreme orbit ensures nearly complete stability of $CH_4$ and therefore even more time for irradiation products to build up.  Sedna also spends most of its orbit beyond the heliopause and is subject to a more intense irradiation environment, with more energetic galactic cosmic rays, than the other two bodies.  This qualitative scenario is consistent with the observations presented here and spectral differences of these three bodies from smaller KBOs.  Vapor pressures of ethane, acetylene, and ethylene are a factor of ~$10^{10}$ or more lower than that of methane at the relevant temperatures, consistent with longer stability (e.g., Fray and Schmitt 2009, Lisse et al. 2021, 2022).  However, detailed modeling of timescales of sublimation from each body over its orbit, irradiation of methane to its immediate byproducts, irradiation of ethane, acetylene, and ethylene to more complex organic materials, impact gardening, sputtering loss, etc will be necessary to quantify the proposed scenario for testing by future observations.

We suggest that Sedna, Gonggong, and Quaoar all underwent internal melting, differentiation, and chemical evolution, similar to the larger dwarf planets and distinct from all smaller KBOs.  The continued presence of light hydrocarbons (ethane, acetylene, ethylene) on the surfaces implies a resupply of methane, the irradiation of which replenishes its immediate irradiation products. By analogy with recently measured isotopic ratios in the ices on Eris and Makemake, it seems likely that the $CH_4$ reservoir present on these bodies to be derived from internal geochemical evolution, highlighting the importance of both size and orbit in determining the present-day surface compositions. The absence of $N_2$ from the surface of the three bodies studied here may be due to its high volatility and early loss, to a smaller original nitrogen reservoir on these bodies than e.g., Pluto, or to differences in geochemical evolution (e.g., lower temperature). Low relative abundances of similarly volatile CO on dwarf planets in general is not well understood, and the lack of strong CO features on these bodies may be due to hydrothermal destruction. The low abundance of $CO_2$ relative to smaller KBOs is consistent with the absence of $CO_2$ absorption features in spectra of Pluto, Eris, and Makemake and may be due to volatile ices and/or irradiation products covering the refractory $CO_2$. A consequence of this scenario is that the surfaces of Sedna, Gonggong, and Quaoar retain very little information about their pre-accretion ingredients, or about their surface composition before migrations put these object in their current orbits, because they are now covered with the irradiation by-products of secondary $CH_4$.

The analysis here focuses on identification of clear absorption features. More insight about these surfaces will be gained by determining relative abundances of materials and their mixing states from future spectral modeling. For instance, there is some indication, based on relative strengths of hydrocarbon features at different wavelengths, for layering. Spectral modeling will also be useful to search for influences of species that we might not have recognized from simple feature identification (e.g., do any N-bearing species affect the shapes of the broader absorption complexes?).

The medium grating spectrum of Quaoar demonstrates the benefit of gaining higher spectral resolution data. The prism spectra of both Sedna and Gonggong contain some weak depressions that are probably unresolved features that would provide even more insight into the surface composition and chemistry. Interpretation of the feature-rich region from 2.1 to 2.7 μm would also benefit from higher spectral resolution, as many narrow bands of various hydrocarbons would remain unresolved in prism data. It will be of particular interest to look at higher resolution for methane bands on Sedna and Gonggong in order to assess the ratio of methane to its irradiation products.

## Acknowledgements

This work is based in part on observations made with the NASA/ESA/CSA James Webb Space Telescope. The data are at the MAST archive at STScI, which is operated by AURA, Inc., under NASA contract NAS 5-03127. These observations are associated with GTO programs 1191 (PI J. Stansberry), 1272 (PI D. Hines), and 1273 (PI J. Lunine). We are especially grateful to Tony Roman, Shelly Meyett, and Charles Proffitt at STScI for help with program implementation. J.A.S. acknowledges support through the sabbatical program at STScI, and the generosity of Lowell Observatory lodging in the Tombaugh apartment for 5 months, and NAU for office facilities during that time. N.P.A. acknowledges support from the Small Research Initiative of the Florida State operated by the Florida Space Institute. R.B. acknowledges support from the CNES.